\documentstyle[12pt]{article}
\def\be{\begin{equation}}
\def\ee{\end{equation}}
\def\bea{\begin{eqnarray}}
\def\eea{\end{eqnarray}}

\begin{document}

\begin{center}

{\Large{\bf Gauging the Superstring under the Group $SU(4)$}}

\vskip .5cm
{\large Davoud Kamani}
\vskip .1cm
{\it Faculty of Physics, 
Amirkabir University of Technology (Tehran Polytechnic)\\
P.O.Box: 15875-4413, Tehran, Iran}\\
{\it e-mail: kamani@aut.ac.ir}\\
\end{center}

\begin{abstract}

The superstring theory in the light-cone gauge
admits various gauge symmetries. Therefore,
we gauge the superstring sigma model in the light-cone gauge
under the gauge group $SU(4)$. 
Some properties of the gauged action and the
corresponding current will be studied. Besides, two modified
actions for the superstring will be obtained.

\end{abstract}
\vskip .5cm



\newpage
\section{Introduction}

The gauged actions in the string theory
have been studied from the various point of view. 
The worldsheet gauge fields have been the object of several 
investigations and they can be introduced in the various
string models \cite{1}-\cite{15}. In other words, the worldsheet gauge fields
provide a concise Lagrangian formulation of different
string models. The two-dimensional $SU(N)$ Yang-Mills theory is part of 
these models \cite{16}-\cite{23}. 

Previously we studied the superstring theory in the presence of a 
$U(1)$ worldsheet gauge field \cite{15}. Here our attention is on the
non-Abelian case. For this, 
we consider the superstring sigma model in the light-cone gauge.
This theory admits a global $SO(8) \subset SO(1,9)$ invariance. One can 
obviously consider the embedding $SU(4) \subset SO(8)$. Therefore, we
gauge this theory under the gauge group 
$SU(4)$, which is a subgroup of the transversal $SO(8)$ group. 

We treat the worldsheet gauge field as an independent degree of freedom. 
In addition, we impose the complex structure in the target space.
The string coordinates form a four-component complex field,
which obeys the gauged Klein-Gordon action.
In the same way, the fermionic degrees of freedom form 
an eight-component complex spinor field. This extended
spinor field obeys the Dirac action with special operator.
However, we shall study some properties of the gauged action and
the current associated to the gauge symmetry. We use this
gauging to modify the superstring action. Therefore, we obtain a 
modified action and an effective action for the superstring.

This paper is organized as follows.
In section 2, the superstring action in terms of extended variables,
appropriate for the $SU(4)$ gauge symmetry,
will be given. In section 3, this action under the
group $SU(4)$ will be gauged. In section 4, two modified actions for the 
superstring will be obtained. Section 5 is devoted to the conclusions.
\section{The superstring action in terms of the complex fields}

The superstring theory in the light-cone gauge is described by 
8 scalars (describing the spacetime coordinates) and 8 Majorana spinors
(describing their fermionic partners) on the worldsheet. The corresponding
action is
\bea
S=-\frac{1}{4\pi \alpha'}\int
d^2\sigma (\eta^{\alpha \beta}\partial_\alpha X^I\partial_\beta X^I
-i{\bar \psi}^I \rho^\alpha \partial_\alpha \psi^I),
\eea
where $I \in \{1,2,...,8\}$ and $\alpha , \beta \in \{0,1\}$. 
The worldsheet metric is
$\eta_{\alpha \beta}=diag (-1,1)$. In the Majorana basis,
the matrices $\rho^0$ and $\rho^1$ are
\bea
\rho^0 = \left( \begin{array}{cc}
0 & -i \\
i & 0
\end{array} \right)\;\;\;,\;\;\;
\rho^1 = \left( \begin{array}{cc}
0 & i \\
i & 0
\end{array} \right).
\eea
This action obviously has the global $SO(8)$ symmetry.

We build the following complex variable from the string coordinates
\bea
Y =\left( \begin{array}{c}
X^1 +i X^2 \\
X^3 +i X^4 \\
X^5 +i X^6 \\
X^7 +i X^8
\end{array} \right).
\eea
Let $\psi^I= \left( \begin{array}{c}
\psi^I_- \\
\psi^I_+
\end{array} \right)$ denote a spinor field of the worldsheet.
Thus, in the same way, these fermionic fields also define the variables
\bea
\Psi_\pm =\left( \begin{array}{c}
\psi^1_\pm +i \psi^2_\pm \\
\psi^3_\pm +i \psi^4_\pm \\
\psi^5_\pm +i \psi^6_\pm \\
\psi^7_\pm +i \psi^8_\pm
\end{array} \right).
\eea
In terms of these extended variables, the action (1) takes the form
\bea
S'=-\frac{1}{4\pi \alpha'}\int
d^2\sigma [\partial_\alpha Y^\dagger\partial^\alpha Y
-2i (\Psi^\dagger_- \partial_+ \Psi_- +\Psi^\dagger_+ \partial_- \Psi_+)],
\eea
where $\partial_\pm = \frac{1}{2} (\partial_\tau \pm \partial_\sigma)$.
This form of the action enables us to study its $SU(4)$ gauge symmetry.

Consider the matrix $U\in SU(4)$ 
and let $U$ be independent of the worldsheet
coordinates $\sigma$ and $\tau$.
Therefore, under the global $SU(4)$ transformations
\bea
&~& Y \longrightarrow UY,
\nonumber\\
&~& \Psi_- \longrightarrow U\Psi_- ,
\nonumber\\
&~& \Psi_+ \longrightarrow U\Psi_+ ,
\eea
the action $S'$ is invariant.
\section{The gauged action}

Now we introduce local vector gauge symmetry by adding a two-dimensional gauge
field as worldsheet dynamical degree of freedom. In other words, we
deform the action (5) to obtain the local $SU(4)$ gauge symmetry.
Thus, we consider the transformations (6) with the coordinate-dependent $U$,
\bea
U(\sigma, \tau)=e^{i\lambda_a (\sigma, \tau) T^a},
\eea
where $\{T^a| a=1, 2, ..., 15\}$ are generators 
and $\{\lambda_a (\sigma, \tau)| a=1, 2, ..., 15\}$ 
are local parameters of the group $SU(4)$.
This implies replacement of the derivative
$\partial_\alpha$ with the covariant derivative $D_\alpha$,
\bea
\partial_\alpha  \longrightarrow D_\alpha = I_{4 \times 4}\partial_\alpha
+ig A_\alpha=I_{4 \times 4} \partial_\alpha +ig A^a_\alpha T^a.
\eea
The gauge field $A_\alpha$ is a $4 \times 4$ Hermitean matrix which lives
in the string worldsheet, and $g$ is the corresponding coupling constant.
According to (8) we should also do the following replacements
\bea
\partial_\pm  \longrightarrow D_\pm = I_{4 \times 4}\partial_\pm
+ig A_\pm,
\eea
where $A_\pm = \frac{1}{2}(A_0 \pm A_1)$.

Adding all these together, we obtain the gauged action
\bea
S_{\rm gauged}=-\frac{1}{4\pi \alpha'}\int
d^2\sigma [(D_\alpha Y)^\dagger D^\alpha Y
-2i (\Psi^\dagger_- D_+ \Psi_- +\Psi^\dagger_+ D_- \Psi_+)
+\pi \alpha'F^a_{\alpha \beta}F^{ \alpha \beta}_a ].
\eea
The field strength of $A_\alpha^a$ is 
\bea
F^a_{\alpha \beta} = \partial_\alpha A_\beta^a - \partial_\beta A_\alpha^a
-g f^a_{\;\;\;bc}A_\alpha^b A_\beta^c,
\eea
where $\{f^a_{\;\;\;bc}\}$ are structure 
constants of the Lie algebra corresponding to the group $SU(4)$.
Since the worldsheet gauge field $A_\alpha$ has been treated as an 
independent degree of freedom ($i.e.$, it is not pull-back of a spacetime
gauge field on the worldsheet) we introduced its kinetic term in (10).
The action (10) under the transformations (6) 
with the local $SU(4)$ matrix (7), and the gauge transformation
\bea
A^a_\alpha \longrightarrow A^a_\alpha  
-\frac{1}{g} \partial_\alpha \lambda^a
-f^a_{\;\;\;bc} \lambda^b A^c_\alpha,
\eea
is symmetric. 
\subsection{A new form for the gauged action}

Define the 8-component complex spinor $\Psi$ as in the following
\bea
\Psi= \left( \begin{array}{c}
\Psi_- \\
\Psi_+
\end{array} \right) .
\eea
Thus, we obtain
\bea
2(\Psi^\dagger_- A_+ \Psi_- +\Psi^\dagger_+ A_- \Psi_+)
={\bar \Psi}\rho^\alpha \otimes A_\alpha \Psi,
\eea
where ${\bar \Psi}$ is defined by
\bea
{\bar \Psi}=\Psi^\dagger \rho^0 \otimes I_{4\times 4},
\eea
in which $I_{4\times 4}$ is the $4 \times 4$ unit matrix.
In the same way, the other fermionic terms in (10) also can be written as
\bea
2 (\Psi^\dagger_- \partial_+ \Psi_- +\Psi^\dagger_+ \partial_- \Psi_+)
={\bar \Psi}\rho^\alpha \otimes I_{4\times 4}\partial_\alpha \Psi.
\eea
In our notation, the direct product of any two matrices $P_{2 \times 2}$ 
and $Q_{4 \times 4}$ is defined by
\bea
P= \left( \begin{array}{cc}
p_1 & p_2 \\
p_3 & p_4
\end{array} \right) \;\;\;,\;\;\;
P \otimes Q = \left( \begin{array}{cc}
p_1 Q & p_2 Q\\
p_3 Q & p_4 Q
\end{array} \right),
\eea
which is a $8 \times 8$ matrix.

Introducing Eqs. (14) and (16) 
into the action (10) gives an elegant
form for the gauged action
\bea
S_{\rm gauged}=-\frac{1}{4\pi \alpha'}\int
d^2\sigma [(D_\alpha Y)^\dagger D^\alpha Y
-i {\bar \Psi}\rho^\alpha \otimes D_\alpha \Psi 
+\pi \alpha' F^a_{\alpha \beta}F^{ \alpha \beta}_a ].
\eea
The eight Majorana spinors of the worldsheet form one Dirac spinor with
eight components. The action of this spinor 
has the feature of the Dirac action with the special
operator $\rho^\alpha \otimes D_\alpha$.
\subsection{The equations of motion}

The equation of motion of the extended variable 
$Y$, extracted from the action (18), is
\bea
D_\alpha D^\alpha Y=0.
\eea
According to the definition of $Y$, 
this equation leads to the equations of motion of 
the string coordinates $\{X^I(\sigma, \tau)\}$. 

For the worldsheet gauge field $A_\alpha$ the equation of motion is
\bea
(D_\beta F^{\alpha \beta})_a
= \frac{g}{4\pi \alpha'}(i [Y^\dagger T_a D^\alpha Y 
- (D^\alpha Y)^\dagger T_a Y ]
-{\bar \Psi}\rho^\alpha \otimes T_a \Psi ).
\eea
For a field which transforms as the adjoint
representation of $SU(4)$ the covariant derivative is
$D_\alpha = \partial_\alpha + ig [A_\alpha ,\;\;]$.
Therefore, the left-hand-side of (20) is
\bea
(D_\beta F^{\alpha \beta})_a =
\partial_\beta F^{\alpha \beta}_a - g f_{abc} A^b_\beta F^{\alpha \beta}_c.
\eea

Finally, the equation of motion of the extended fermionic field $\Psi$ is
\bea
\rho^\alpha \otimes D_\alpha \Psi =0.
\eea
This equation is decomposed into the equations of motion of 
the components of the worldsheet fermions, $i.e.$ 
$\{\psi^I_\pm (\sigma , \tau)\}$.
\subsection{The corresponding current}

We have the transformations (6) with the local matrix (7), and also 
the transformation (12). The fermionic parts of (6) can be written as
\bea
\Psi \longrightarrow I_{2 \times 2} 
\otimes e^{i\lambda_a (\sigma, \tau) T^a}
\Psi .
\eea
Thus, the infinitesimal transformations of $Y$, $\Psi$ and $A^a_\alpha$ are
\bea
&~& \delta Y = i \lambda_a T^a Y ,
\nonumber\\
&~& \delta \Psi = i \lambda_a I_{2 \times 2} \otimes T^a \Psi ,
\nonumber\\
&~& \delta A^a_\alpha =-\frac{1}{g} (D_\alpha \lambda)^a
= -\frac{1}{g} \partial_\alpha \lambda^a -f^a_{\;\;\;bc} 
\lambda^b A^c_\alpha\;.
\eea
Introducing these transformations into the variation of the action (18)
gives the current 
\bea
J^\alpha_a = \frac{g}{4\pi \alpha'}(i [Y^\dagger T_a D^\alpha Y 
- (D^\alpha Y)^\dagger T_a Y ]
-{\bar \Psi}\rho^\alpha \otimes T_a \Psi ).
\eea
This is a Hermitean current.

According to the current (25), Eq. (20) takes the form
$(D_\beta F^{\alpha \beta})_a=J^\alpha_a$. 
This implies that the current $J^\alpha=J^\alpha_a T^a$
transforms covariantly under the gauge group $SU(4)$.
Therefore, by the equations of motion, 
it is a covariantly constant quantity
\bea
D_\alpha J^\alpha =\partial_\alpha J^\alpha + ig [A_\alpha , J^\alpha]=0.
\eea

\subsubsection{The current due to the global gauge symmetry}

The current (25) can be decomposed as in the following
\bea
J^\alpha_a = J^{(0)\alpha}_a -\frac{g^2}{4\pi \alpha'} A^\alpha_b Y^\dagger
\{T_a, T_b\} Y,
\eea
where
\bea
J^{(0)\alpha}_a = \frac{g}{4\pi \alpha'}
[i (Y^\dagger T_a \partial^\alpha Y
- \partial^\alpha Y^\dagger T_a Y ) 
- {\bar \Psi}\rho^\alpha \otimes T_a \Psi ],
\eea
is the current $J^\alpha_a$ in terms of $A_\alpha=0$.

Now consider the action (18) with $A_\alpha=0$,
\bea
S_{\rm un-gauged}=-\frac{1}{4\pi \alpha'}\int
d^2\sigma (\partial_\alpha Y^\dagger \partial^\alpha Y
-i {\bar \Psi}\rho^\alpha \otimes I_{4\times 4}\partial_\alpha \Psi).
\eea
This is the un-gauged action (5), which is symmetric under the global
$SU(4)$ transformations (6).
The current, associated to this 
global symmetry, is given by $J^{(0)\alpha}_a$.
In this case the conservation law is $\partial_\alpha J^{(0)\alpha}_a =0$.
For showing this, use the equations of motion (19) and (22) with
$A_\alpha=0$, and their Hermitean conjugates, $i.e.$ 
$\partial_\alpha \partial^\alpha Y^\dagger=0$ and 
$\partial_\alpha {\bar \Psi} \rho^\alpha \otimes I_{4\times 4}=0$.
In addition, for the fermionic part apply the identities
\bea
\rho^\alpha \otimes T_a 
= (I_{2 \times 2}\otimes T_a)(\rho^\alpha \otimes I_{4\times 4})
=(\rho^\alpha \otimes I_{4\times 4})(I_{2 \times 2}\otimes T_a).
\eea
\section{Gauge modifications of the superstring action}

In this section we impose the assumption that 
the kinetic term of the gauge field, in the gauged action (18),
to be absent. Therefore, we receive the action
\bea
I=S + S_1,
\eea
where
\bea
S_1= -\frac{1}{4\pi \alpha'} \int d^2 \sigma \{ ig [\partial_\alpha
Y^\dagger A^\alpha Y- Y^\dagger A_\alpha \partial^\alpha Y
-i {\bar \Psi}\rho^\alpha \otimes A_\alpha \Psi ]
+ g^2 Y^\dagger A_\alpha A^\alpha Y \}. 
\eea
This implies that $A_\alpha$ is an auxiliary field. 
\subsection{A modified action}

The equation of motion of $A_\alpha$ has been given 
by (20) without the left-hand-side.
Thus, we obtain the gauge field as in the following
\bea
A^\alpha_a = \frac{4\pi \alpha'}{g^2}(M^{-1})_{ab}J^{(0)\alpha}_b ,
\eea
where the symmetric matrix $M$ is defined by the elements
\bea
M_{ab}= Y^\dagger \{T_a , T_b\}Y.
\eea
The equation (33) reveals an explicit relation between string and 
the $SU(4)$ gauge field.

According to (33) the action (32) can be written as
\bea
S'_1={\rm Tr} \int d^2 \sigma A_\alpha J^\alpha_{(0)}.
\eea
This implies that $J^\alpha_{(0)}$ acts as an external current 
source for the $SU(4)$ gauge field $A_\alpha$.
Combining (33) and (35) gives
\bea
S''_1=\frac{1}{8\pi \alpha'} \int d^2 \sigma {\cal{J}}^\alpha_a
(M^{-1})_{ab} {\cal{J}}_{\alpha_b},
\eea
where ${\cal{J}}^\alpha_a=\frac{4\pi \alpha'}{g}J^{(0)\alpha}_a$. 
This action is independent of the gauge coupling constant $g$.
Finally, we obtain the action
\bea
I_{\rm Modified}= S+S''_1.
\eea
In fact, this demonstrates a gauge modification for the superstring action,
originated from the $SU(4)$-gauging.
\subsection{An effective action}

By path integration over $A^a_\alpha$ we obtain an effective action of $I$,
\bea
e^{-i I_{\rm eff}}=\int \Pi^{15}_{a=1}\Pi^1_{\alpha=0} D A^a_\alpha e^{-iI}.
\eea
For computing $I_{\rm eff}$, we should introduce the Wick's rotation
$\tau \rightarrow it$. After calculations and imposing inverse of the
Wick's rotation we receive the effective action as in the following
\bea
I_{\rm eff}=S+S''_1+S_2+C,
\eea
where $S_2$ is 
\bea
S_2=-\frac{1}{2} {\rm Tr}\int d^2 \sigma \ln M.
\eea
The constant $C$ also is
\bea
C=-\frac{15}{2}V_\Sigma \ln \bigg{(} \frac{g^2}{4\pi\alpha'}\bigg{)},
\eea
in which $V_\Sigma$ is the area of the string worldsheet. In fact, 
this effective action is another gauge modification of the superstring
action, due to the $SU(4)$ gauge field.
\section{Conclusions}

The superstring sigma model in the light-cone gauge  
enabled us to gauge it under the group $SU(4)$. 
We observed that this action can be written in terms of the four-component
and eight-component
complex variables, which are constructed from the worldsheet fields. 

The fermionic part of the gauged action
has the feature of the Dirac action with 
an eight-component complex  spinor field
and a special matrix-derivative operator. 
The $SU(4)$ worldsheet gauge field couples to this spinor field.

We demonstrated that this model contains some connections between
string and the $SU(4)$ gauge field. This enabled us to obtain a modified
action for string, due to the $SU(4)$-gauging. Similarly, removing the
gauge field through the path integral gave an effective action for
the string.

In a similar fashion one can combine the worldsheet fields to obtain the
quaternionic and or octanionic variables. The former enables us to gauge 
the action under the group $SU(2)$ and the 
latter provide gauging under the group $U(1)$.

Note that consideration of two-dimensional gauge field induces a contribution
to the conformal anomaly and hence changes the critical 
dimension of the string models.

\end{document}